# Novel Rotor Fault Diagnostic Method Based on RLMD and HT Techniques

Asma Guedidi[1] and Widad laala[2]

[1]Department of electrical Engineering, Mohamed khider
biskra University, Biskra, Algeria
[2]Department of electrical Engineering, Mohamed khider biskra University,
Biskra, Laboratory of electrical engineering Biskra (LGEB)

## Abstract

*Frequency domain analysis using the Fast Fourier transform (FFT) has been a popular method for diagnosing broken rotor bar (BRB) faults in squirrel-cage induction motors (IM). However, FFT analysis is limited by sampling frequency and time acquisition constraints, making it less effective under time-varying conditions. To overcome these difficulties, a novel BRB fault detection method for non-stationary conditions is proposed. The proposed strategy is based on the recently developed robust local mean decomposition (RLMD) and Hilbert transform (HT) methods. Using these techniques, the BRB characteristic frequency and amplitude component are obtained from only one phase stator current allowing automation of the features detection process. in fact, HT is used to extract the stator current envelope (SCE). Then, the SCE is processed by RLMD for determining the sub signals production functions (PFs). Finally, HT is applied to the most sensible PF to compute its instantaneous frequency and amplitude. The tracking of the BRB fault characteristic can inform us about the condition of the induction motor. The effectiveness of the proposed diagnostic strategy is validated through simulation conducted in the Matlab environment. The simulation results show the capability of this method to track accurately the frequency and amplitude of the 2sf component where f and s represent the fundamental stator current frequency and motor slip respectively*

## 1. Introduction

Induction motors are still among the most reliable and important electrical machines. The wide range of their use involves various electrical, magnetic, thermal and mechanical stresses which results in the need for fault diagnosis as part of the maintenance [1].

The Broken Rotor Bar (BRB) fault is one of the predominant failure modes of induction machines, The consequences of this fault include excessive vibrations, poor starting performances, torque fluctuation, and high thermal stress. If this fault remains undetected it may lead to potentially catastrophic failures. Thus, it is important to detect this particular fault to prevent permanent failure of induction machine [2].

The monitoring and diagnosis of induction machines based on stator-current signature analysis (MCSA) are commonly used as a noninvasive method for detection of electrical as well as mechanical faults, this method utilizes the machine stator current frequency signatures to detect the BRB fault. However, Although MCSA based on fast Fourier transform (FFT) is a great method and is the most widely used technique for fault detection in IM due to its high accuracy and simplicity, but its application is limited due to some inherent drawbacks. In fact, the MCSA method is based on analyzing the signals only in the frequency domain which mean, it does not provide the time information which is very important in many applications. Moreover, an accurate spectrum analysis of stator current has been achieved by obtaining a large number of data points. However, it is not always possible to take a large number of data points due to the limitation of the digital system memory size [3]. Furthermore, FFT approach is not suitable for the non- stationary conditions which therefore limit its use. Inspired from this, other alternative methods were introduced based on time-frequency analysis (TFA). Indeed, Time–frequency analysis can identify the signal frequency components, reveals their time variant features, and is an effective tool to extract machinery health information contained in nonstationary signals. In fact, the basic goal of the TFA is to determine the energy concertation along the frequency axis at a given time instant, to search for joint time-frequency representation of the signal. [4]. Among its common technique There are short time Fourier transform (STFT); Wavelet transform (WT), empirical mode decomposition (EMD) and variational mode decomposition (VMD).

Technically, Wavelet transform (WT) is an improved method in term to address the shortcoming of short STFT which is based on a fixed window size that maintain a constant resolution in both time and frequency domain unlike the wavelet transform who is developed with many various windows which allow a multi-scale time- frequency resolution of the signal but how to choose the proper wavelet basis function is a difficult task. Empirical mode decomposition (EMD) is an entirely data-driven and highly adaptive signal analysis method. However, recent study shows that this method suffers from mode mixing problem which makes a lot of EMD-improvised version have been proposed like ensemble EMD (EEMD). However, as noise- aided algorithm EEMD will inevitably leave residual white noise in the decomposed signal. In addition, when the amount of white noise added is large, the average number of integrations also needs to be increased, which results in taking more time to decompose the signal [4]. Variational mode decomposition is proposed as an alternative method to the EMD in term of optimizing the decomposition modes. Nevertheless, its performance is strictly limited by the appropriate selection of the balancing parameter of the data fidelity constraint (alfa) and the number of components (k) to be extracted. Another new adaptive analysis technique called local mean decomposition (L.M.D) was initially developed by Smith [5]. LMD decomposes the complex non-stationary data into a number of product functions (PFs) [6], [7] [8]. Each PF is a product between the signal envelope and the frequency-modulated (FM) signal which is produced by varying the wave instantaneous frequency. In contrast to EMD, LMD has a low iterative decomposition rate and can immediately produce the instantaneous frequency (FM) and the instantaneous amplitude (AM) without Hilbert transform (HT) [6], [7] [8]. However, LMD also suffers from the same limits of EMD, the mode mixing and the end effect [7], [8]. To address this situation, Liu et al. [6], [8] have developed the conventional LMD by adjusting an appropriate parameter selection for boundary condition, envelope estimation and sifting stop criterion, all related with end effect and mode mixing issues [7], [8]. Most researchers have ignored that these three parameters are not independent of each other and focused on only one parameter at a time [8], [6],[7]. Therefore, a robust optimization approach, which is named by the author "robust LMD" (RLMD), includes the three algorithms to improve the LMD performance through an integrated parameter selection process [6], [7]. The first algorithm is a mirror extending approach to deal with the boundary condition [6], [8]. The second algorithm is a moving average which is employed as smooth algorithm for envelope estimation for both local mean and local magnitude in LMD in addition to a self-adaptive tool based on statistics theory to automatically calculate the fixed subset size of moving average for accurate envelope estimation [8], [7]. Finally, a soft sifting stopping criterion is suggested to allow LMD to accomplish a self-adaptive stop in every sifting process [5], [8]. In the final phase, an objective function takes into account the global and local characteristics of the target signal. Based on this function, a heuristic mechanism is provided to automatically define the optimum sifting iterations number in the sifting process [5], [6].

This paper introduces a novel diagnostic method for the early detection and characterization of broken rotor bar (BRB) defects. The main contribution of the proposed methods lies time-frequency representations, which are effective for fault diagnosis in induction motors. These methods aim to extract valuable information to detect and identify fault indicator frequencies. The novelty of this paper lies in monitoring the amplitude and frequency of 2gf BRB component by applying the combined HT and RLMD to the stator current. The diagnostic method has been validated through simulations under different fault severity and load conditions.

Therefore, this paper is summarized as follows, the first section presents an introduction, the second section present a brief introduction of the proposed methods, the reminder section represents the application of the proposed methods and its results. Finally, Section fourth concludes this work.

## 2. THEORICAL BACKGROUND

### 2.1. Hilbert Transform Technique

HT is a well-known signal analysis method. It is based on exploiting the content of the envelope of the current signal which calculated down below:

The HT of a real signal $i_A(t)$ is defined as [10]:

$$HT(i_A(t)) = y(t) = \frac{1}{\pi t * i_A(t)} = \frac{1}{\pi} i_A(\tau) d\tau / (t - \pi)$$

The so-called analytic signal $i(t)$ combination of the real signal with its HT:

$$i(t) = i_A(t) + jy(t) = a(t) * \exp(k\phi(t))$$

Where:

$$a(t) = \sqrt{i_a(t)^2 + y(t)^2}$$

$$\varphi(t) = \arctan(y(t)/i_A(t))$$

A(t) is the instantaneous amplitude of i(t) known as the envelope of the $i_A(t)$ and $\varphi(t)$ is the instantaneous phase of i(t)

### 2.2. Robust Local Mean Decomposition

The RLMD method enhances LMD by addressing three key areas, effectively mitigating issues related to modal mixing and endpoint effects. The specific optimizations for each aspect are as

follows [11], [12]:

Step 1 (Optimization of Boundary Conditions): The boundary conditions are optimized using the mirror extension method for signal extension, a technique commonly employed in various modal decomposition methods.

Step 2 (Optimization of Envelope Estimation of the Signal): First, determine the step size of the signals $m_0(n)$ and $a_0(n)$ using the values is $e_{k+1} - e_k + 1$. Next, apply a histogram meta-count to the step set to obtain the probabilities $S(k)$ and $edge(k)$ for each meta, defining the center of the step $\mu_s$, and the standard deviation $\delta_s$ as per Equation (1) and (2). Finally, calculate the size of the selected subset as $k = odd\left(\mu_s + 3\delta_s\right)$

$$\mu_s = \sum_{k=1}^{N_b} s(k)S(k)$$

$$\delta_s = \sqrt{\sum_{k=1}^{k}[s(K) - \mu_s]^2 S(k)}$$

Where $s(k) = [edge(k) + edge(k+1)]/2$

Step 3 (Optimization of Sifting Stopping Criteria): $a_{ij}(n)$ is the smoothed local amplitude generated by the $j-th$ iteration during the sifting process for the $i-th$ PF component. In each iteration, the objective function is defined by Equation (3). If the process $f_{ij+1} > f_{ij}$ and $f_{ij+2} > f_{ij+1}$ stops at the $j-1-th$ iteration and the corresponding result is returned; otherwise, the iteration continues until the maximum number of iterations is reached.

$$f = RMS(z(n)) + EK(z(n))$$

Where RMS denoted to root mean square, EK denote to excess kurtosis and Zero baseline envelope signal is $z(n) = a(n) - 1$

## 2.3. Stator Phase Current in Case of Rotor Fault

Theoretically, in the case of rotor asymmetry created by broken bars, the stator current can be written as:

$$I_A(t) = I_f \cos(\varphi) + \sum_k ((I_{rbb1}^k \cos\varphi_{rbb1}^k + I_{rbb2}^k \cos\varphi_{rbb2}^k)\cos(2\pi(2ksf)t)$$
$$+ (I_{rbb2}^k \sin\varphi_{rbb2}^k - I_{rbb1}^k \sin\varphi_{rbb1}^k)\sin(2\pi(2ksf)t))) \quad (11)$$

where: $I_f$ The fundamental value of the phase stator current. $\varphi$ The main phase shift angle of the phase stator current. $I_{rbb1}^k$ The left magnitude for the harmonic component $f_{rbb}$ $I_{rbb2}^k$, The right magnitude for the harmonic component $f_{rbb}$. $\varphi_{rbb1}^k$, The left phase shift angle of component $f_{rbb}$. $\varphi_{rbb2}^k$, The right phase shift angle of component $f_{rbb}$.

the Expression (1) can be rewritten as:

$$i_A(t) = A(t)\cos(2\pi ft) + B(t)\sin(2\pi ft) \qquad (12)$$

12) can take the following form

$$i_A(t) = A_m(t)\sin(2\pi ft + \theta(t)) \qquad (13)$$

With

$$A_m(t) = \sqrt{A(t)^2 + B(t)^2} \qquad (14)$$

$$\theta(t) = \arctan(A(t)/B(t)) \qquad (15)$$

$$A(t) = I_f \cos(\varphi) + \sum_k ((I_{rbb1}^k \cos\varphi_{rbb1}^k + I_{rbb2}^k \cos\varphi_{rbb2}^k)\cos(2\pi(2ksf)t$$
$$+ (I_{rbb2}^k \sin\varphi_{rbb2}^k - I_{rbb1}^k \sin\varphi_{rbb1}^k)\sin(2\pi(2ksf)t))) \qquad (16)$$

$$B(t) = I_f \sin(\varphi) + \sum_k ((I_{rbb1}^k \sin\varphi_{rbb1}^k + I_{rbb2}^k \sin_{rbb2}^k)\cos(2\pi(2ksf)t$$
$$+ (I_{rbb2}^k \cos\varphi_{rbb2}^k - I_{rbb1}^k \cos\varphi_{rbb1}^k)\sin(2\pi(2ksf)t))) \qquad (17)$$

As shown in previous relations (16) and (17), the rotor asymmetry faults in induction motor induced by the broken rotor bars modulate the stator current amplitude at frequency 2ksf, by exploiting this fact; the stator current envelope can be used as diagnostic signal.

## 3. SIMULATION RESULTS

The mathematical model used to generate the stator current is well described in [9]. The motor is a star-connected 3 kW machine, with four poles and 28 bars. The motor was supplied by a three-phase, 50 Hz, sinusoidal balanced supply. At the rated voltage amplitude of 380 V. The current signal is sampled at the frequency of 10 kHz. The proposed algorithms were implemented in MATLAB and tested using simulated signals. Our goal is to efficiently extract the instantaneous characteristics (amplitude/frequency) of the fault from the stator envelope current (SCE) signal under variable conditions. Fig.1.a illustrates the simulated stator current(drawn in blue) and its envelope (drawn in red) for healthy state. Whereas, Fig.1.b shows the same signals with the fault of two broken bars. Followed by a sudden increase in the fault severity, it is clearly shown that the amplitude modulation of stator current was affected by the rotor asymmetry due to the BRB fault, this effect is embedded in the envelope signal where the currents in the faulted motors exhibit significant fluctuations. However, the aim of this work is to extract and exploit this information through signal processing and diagnostic methods. Thus, we will apply the first algorithm which is Hilbert transform technique to extract the stator current envelope (SCE). Then, the SCE was decomposed by RLMD into a set of PFs as shown in (Fig. 2), the amplitude and the frequency related of each component under non stationary conditions are illustrated respectively in Figs 3 and 4. According to these results, we can notice that the instantaneous amplitude of the first component react with the fault evolution. in the other hand, the time-frequency figure gives an ambiguous information. To address this issue, Hilbert transform is applied to each component. The analysis of the first component, presented in Figs 5 and 6, show that the severity defect introduces significant distortions. after t = 0.9 s, the SCE fluctuation amplitude which represent the 2sf amplitude increase significantly. However, we notice that the frequency value 2sf (about 6 Hz) fault harmonic remains unchangeable due to the load invariance (s = 0.06).

In order to demonstrate the tracking capabilities of the proposed diagnostic method, a simulation analysis was performed, in which a sudden load variation (from 15 Nm to 20 Nm) was introduced during the stator current recording period, for a motor with two broken bars. The time waveform of the stator current and the SCE are presented in Fig. 7. The obtained PFs after the decomposition of SCE via RLMD is given in fig 8. In this event, the instantaneous amplitude of the first PF increases proportionally to the load (Fig 9). However, its instantaneous frequency (IF) fails to track the correct value of the 2sf fault frequency component (Fig 10). For this reason, the association of HT with RLMD perform the accuracy of tracking the amplitude and frequency process as shown in Figs 11 and 12.

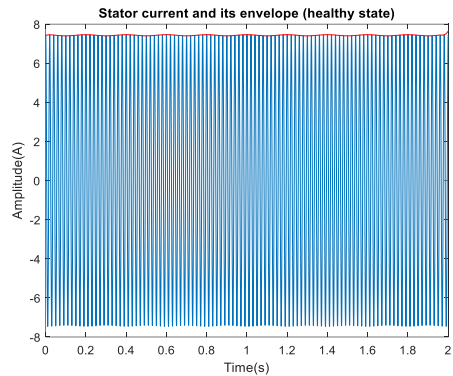

(a)

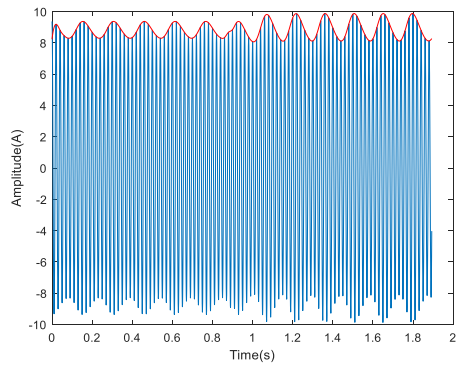

(b)

Fig.1 the stator current and its envelope: a) healthy state b) with increase of fault severity

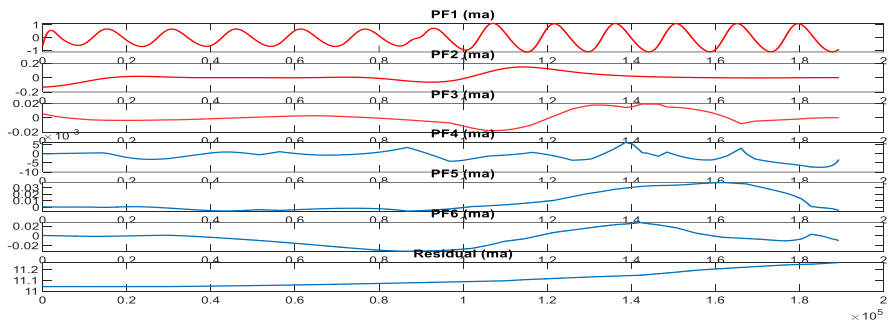

Fig.2 The decomposed sub-signals PFs by the application of RLMD to the envelope current signal

Computer Science & Information Technology (CS & IT) 23

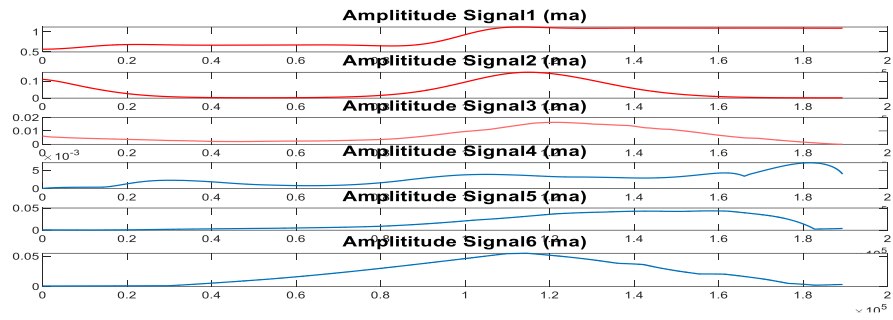

Fig.3 The instantaneous amplitude of each PFs obtained by RLMD

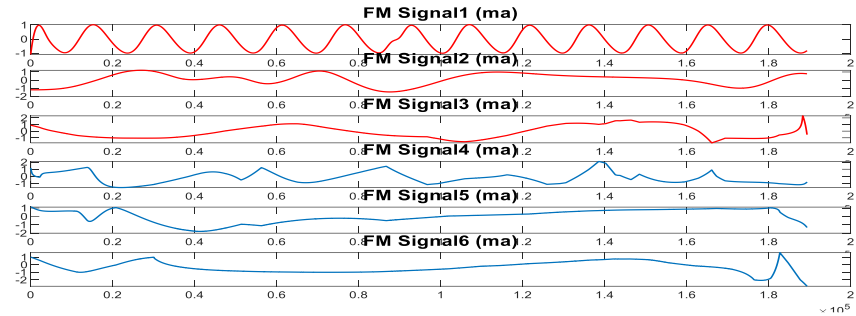

Fig.4 The instantaneous frequency of each PFs obtained by RLMD

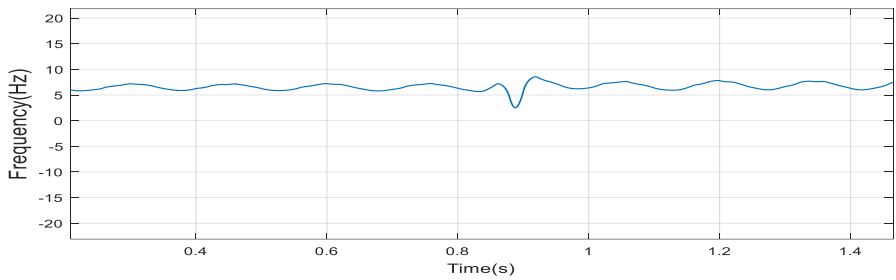

Fig.5 The instantaneous frequency of dominant PF obtained by HT

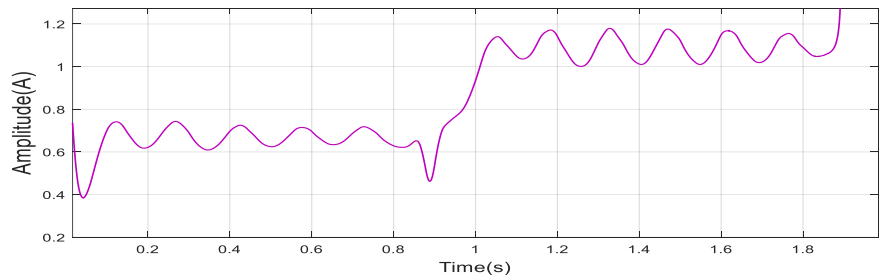

Fig.6 The instantaneous amplitude of dominant PF obtained by HT

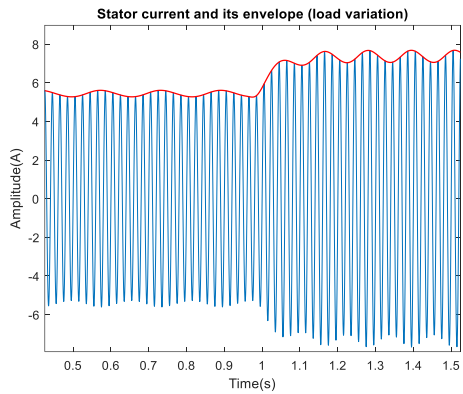

Fig.7 The stator current and its envelope of faulty motor with load variation suddenly at t=1s

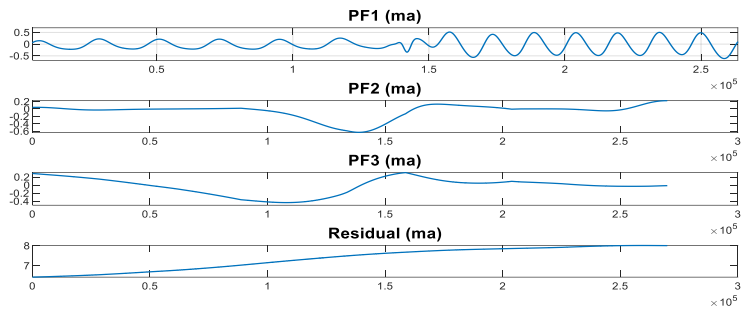

Fig.8 The decomposed sub-signals PFs obtained by the application of RLMD to the envelope current signal

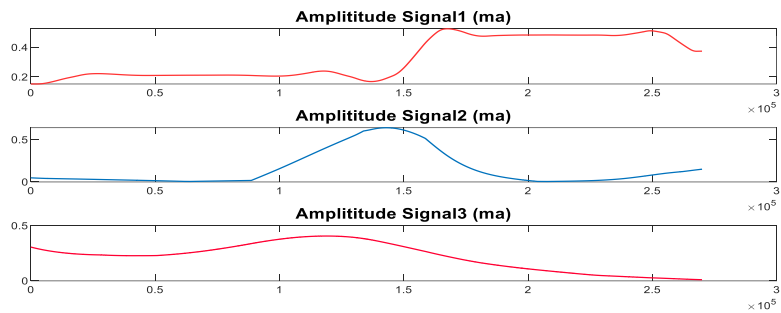

Fig.9 The instantaneous amplitude of each PFs obtained by RLMD

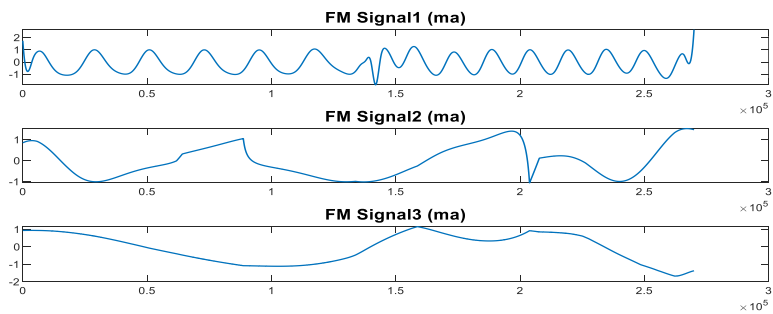

Fig.10 The instantaneous frequency of each PFs obtained by RLMD

Computer Science & Information Technology (CS & IT) 25

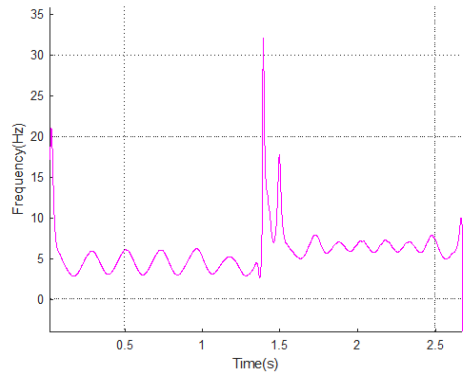

Fig.11 The instantaneous frequency of dominant PF obtained by HT

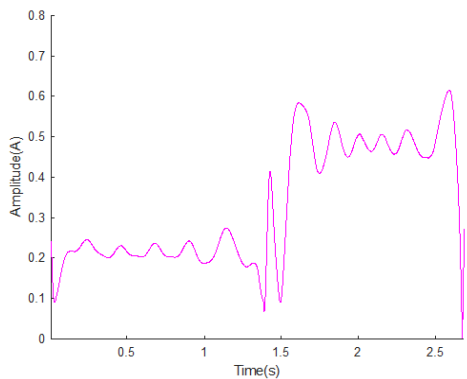

Fig.12 The instantaneous amplitude of dominant PF obtained by HT

## 4. CONCLUSION

This paper investigates a novel method for diagnosing broken rotor bar faults under variable conditions by analyzing the stator current of a three-phase induction motor. The proposed approach combines the enhanced Local Mean Decomposition (RLMD) with the Hilbert Transform (HT) to track the instantaneous frequency and amplitude of the faulty component in non-stationary conditions. The HT method is used to extract the stator current envelope (SCE), which contains significant local features, thereby meaningfully reducing complex numerical calculations. Then, the SCE is processed via RLMD and HT for identifying the BRB faulty components. **The study results .**demonstrate the ability of the proposed diagnostic strategy to evaluate the fault severity in detecting the BRB faults by tracking the instantaneous amplitude and frequency of 2fs BRB component.